\documentclass[aps,preprint, showpacs]{revtex4}%
\usepackage{amsfonts}
\usepackage{amsmath}
\usepackage{amssymb}
\usepackage{graphicx}%
\usepackage{hyperref}
\begin{document}
\title{Naked and truly naked rotating black holes}

\author{H. V. Ovcharenko}
\email{hryhorii.ovcharenko@matfyz.cuni.cz}
\affiliation{Department of Physics, V.N.Karazin Kharkov National University, 61022 Kharkov, Ukraine}
\affiliation{Institute of Theoretical Physics, Faculty of Mathematics and Physics, Charles
University, Prague, V Holesovickach 2, 180 00 Praha 8, Czech Republic}

\author{O. B. Zaslavskii}
\email{zaslav@ukr.net}
\affiliation{Department of Physics and Technology, Kharkov V.N. Karazin National
University, 4 Svoboda Square, Kharkov 61022, Ukraine}

\begin{abstract}
Previously, it was noticed that in some space-times with Killing horizons some
curvature components, responsible for tidal forces, small or even zero in the
static frame, become enhanced from the viewpoint of a falling observer. This
leads to the notion of so-called naked black holes. If some components in the
frame attached to a free-falling observer formally diverge, although scalar
invariants remain finite, such space-times was named "truly naked black holes"
(in mathematical language, one can speak about non-scalar singularity).
Previous results included static spherically symmetric or distorted static
metrics. In the present work, we generalized them to include rotation in
consideration. We also scrutiny how the algebraic type can change in the
vicinity of the horizon due to local Lorentz boost. Our approach essentially
uses the Newman-Penrose formalism, so we analyze the behavior of Weyl scalar
for different kinds of observers.

\end{abstract}

\maketitle
\tableofcontents

\newpage

\section{Introduction}

The main feature typical of a black hole consists in the existence of the
event horizon. This strongly separates two classes of observers: those who
reside outside the horizon and those who fall in the inward direction.
Correspondingly, this makes comparison of properties of both families of
observers especially important. One of questions that arise here is how they
perceive surrounding geometry. It turned out that quite nontrivial situations
may arise here. Some curvature components, small or even zero in the static
frame can become enhanced from the viewpoint of a falling observer \cite{nk1},
\cite{nk2}. This made it quite natural to pose a question: when do some
components of the curvature tensor in the frame attached to a falling observer
become not simply large but formally diverging \cite{pravda2005},
\cite{Zaslavskii2007}? From the mathematical viewpoint, this can be considered
as an example of so-called nonscalar curvature singularity \cite{he}, \cite{e}
that finds realization in physically relevant context within black hole
physics. In particular, this affects the classification of gravitational field
since for different observers a Petrov type may be different \cite{pravda2005}%
, \cite{Tanatarov2014}. Usually, restrictions on the behavior of physical
quantities are formulated in the frame attached to falling observers
\cite{membrane}, so now these conditions can be violated since geometry itself
fails to be regular.

The objects under discussion were named "truly naked black holes" (TNBHs).
Strictly speaking, TNBH is not a black hole at all \cite{kh}. However, for
shortness, we adhere to this (not quite rigorous) term implying that this is
not a black hole which is truly naked but a separate object on its own (so we
consider TNBH as an integral term). In a similar manner, the expression
"singular horizon" is used in literature for shortness in spite of the fact
that it is not a horizon, if it is not regular. More general questions should
include properties of material content and their overlap with properties of
geometry in what concerns regularity of horizons or its violation
\cite{curly}, \cite{inmatter}, \cite{neutral}, \cite{kh}. (According to
terminology of \cite{visser}, we call black holes dirty if they are surrounded
by matter.). Thus we extend considerations of \cite{Tanatarov2014},
\cite{vis04prd} where only regular space-times were considered.

Investigation of general conditions of regularity and properties of TNBH was
carried out earlier for static spherically symmetric and distorted black
holes. In the present work we make the next step and consider rotating
stationary space-times. The case of completely regular horizons in such
space-times was studied in \cite{dirty}, \cite{Ovcharenko2023}. Now, we look
which and how some conditions of regularity can be relaxed realized to admit
TNBH. With this reservation in mind, we exploit and extend previous
classification. We called a black hole usual if these tidal forces tend to
zero in the frame comoving with the observer, naked if they are finite, and
truly naked if they diverge \cite{Zaslavskii2007}, \cite{hor}. Now, this
classification will be generalized and reformulated in the terms of Weyl
scalar to make it convenient for use in the framework of the Newman-Penrose
formalism \cite{np}.

New line of motivation for study of axially symmetric TNBH comes from a recent
work \cite{extr} where it is stated that higher order curvature corrections in
gravitational Lagrangian destroy regular extremal horizons that existed in
general relativity for the Kerr metric, thus making them singular. This poses
a question, which possibilities for regular horizons exist in principle in any
theory. And, if some of regularity conditions fail, how this manifests itself
in the properties of geometry and observers who probe it?

\section{General setup}

We consider axially symmetric spacetimes
\begin{equation}
ds^{2}=-N^{2}dt^{2}+g_{\varphi\varphi}(d\varphi-\omega dt)^{2}+\dfrac{dr^{2}%
}{A}+g_{\theta\theta}d\theta^{2}. \label{metr}%
\end{equation}

Our convention in the choice of coordinates is $x^{0}=t,~x^{1}=r,~x^{2}%
=\theta,~x^{3}=\varphi$. We choose a so-called zero-angular momentum observer
(ZAMO) \cite{bpt} to which we attach the tetrad
\begin{align}
e_{(0)}^{\mu}  &  =\dfrac{1}{N}(1,0,0,\omega)~~~e_{(1)}^{\mu}=\sqrt
{A}(0,1,0,0)\nonumber\\
e_{(2)}^{\mu}  &  =\dfrac{1}{\sqrt{g_{\theta\theta}}}(0,0,1,0)~~~e_{(3)}^{\mu
}=\dfrac{1}{\sqrt{g_{\phi\phi}}}(0,0,0,1) \label{ozamo_fr}%
\end{align}
In addition, we can introduce a null tetrad
\begin{align}
k  &  =\dfrac{e_{(0)}+e_{(1)}}{\sqrt{2}},~~~l=\dfrac{e_{(0)}-e_{(1)}}{\sqrt
{2}}\nonumber\\
m  &  =\dfrac{e_{(2)}+ie_{(3)}}{\sqrt{2}},~~~\bar{m}=\dfrac{e_{(2)}-ie_{(3)}%
}{\sqrt{2}} \label{null_fr}%
\end{align}
We want to define under which conditions a black hole is naked. First of all,
we require that curvature scalars for such black holes be finite on the
horizon. If it is so, then a point-like particle moving under the action of
finite forces can cross the horizon. However, if a particle has small but
nonzero size, one can ask a question about tidal forces acting on a particle
which are related to the Riemann curvature tensor.

Now let us analyze the Riemann tensor. Our aim is to analyze under which
conditions the components of the Riemann tensor are finite in a frame,
comoving with a falling observer. To solve this problem, we choose such a strategy.

\begin{itemize}
\item \textbf{Decompose the Riemann tensor on the Weyl tensor, Ricci tensor,
and Ricci scalar in the tetrad frame (\ref{ozamo_fr})}. This can be done using
a definition of the Weyl tensor:
\begin{align}
C_{(a)(b)(c)(d)}=  &  R_{(a)(b)(c)(d)}-\dfrac{1}{2}\left(  \eta_{(a)(c)}%
R_{(b)(d)}+\eta_{(b)(d)}R_{(a)(c)}-\eta_{(a)(d)}R_{(b)(c)}-\eta_{(b)(c)}%
R_{(a)(d)}\right)  +\nonumber\\
&  +\dfrac{1}{6}R\left(  \eta_{(a)(c)}\eta_{(b)(d)}-\eta_{(a)(d)}\eta
_{(b)(c)}\right)  \label{weyl_orig_tetr}%
\end{align}
where $R_{(a)(b)(c)(d)}=R_{\mu\nu\rho\sigma}e_{(a)}^{\mu}e_{(b)}^{\nu}%
e_{(c)}^{\rho}e_{(d)}^{\sigma}$ is the Riemann tensor in the tetrad frame,
$\eta_{(a)(b)}$ being the Minkowski metric, $R_{(a)(b)}=R_{(a)(c)(b)}^{(c)}$
Ricci tensor, $R=R_{(c)}^{(c)}$ Ricci scalar.

\item \textbf{Introduce Weyl scalars and Ricci tensor components in the null
tetrad (\ref{null_fr})}. Weyl scalars are defined as
\begin{align}
\Psi_{0}  &  =C_{abcd}k^{a}m^{b}k^{c}m^{d},~~~~\Psi_{4}=C_{abcd}l^{a}\bar
{m}^{b}l^{c}\bar{m}^{d},\\
\Psi_{1}  &  =C_{abcd}k^{a}l^{b}k^{c}m^{d},~~~~~~\Psi_{3}=C_{abcd}k^{a}%
l^{b}\bar{m}^{c}l^{d},\\
\Psi_{2}  &  =C_{abcd}k^{a}m^{b}\overline{m}^{c}l^{d}%
\end{align}
and, similarly, the Ricci tensor components
\begin{align}
\Phi_{00}  &  =\dfrac{1}{2}R_{\mu\nu}k^{\mu}k^{\nu},~~~~~\Phi_{22}=\dfrac
{1}{2}R_{\mu\nu}l^{\mu}l^{\nu},\\
\Phi_{01}  &  =\dfrac{1}{2}R_{\mu\nu}k^{\mu}m^{\nu},~~~~\Phi_{12}=\dfrac{1}%
{2}R_{\mu\nu}l^{\mu}m^{\nu},\\
\Phi_{02}  &  =\dfrac{1}{2}R_{\mu\nu}m^{\mu}m^{\nu},~~~\Phi_{11}=\dfrac{1}%
{4}R_{\mu\nu}\left(  k^{\mu}l^{\nu}+m^{\mu}\bar{m}^{\nu}\right)
\end{align}
Thus, knowing $\Psi_{i}$, $\Phi_{ij}$ and $R$ one can restore the Riemann
tensor, simply inverting (\ref{weyl_orig_tetr}). Moreover, if $\Psi_{i}$,
$\Phi_{ij}$ and $R$ are finite, then the Riemann tensor is also finite.
Explicit expressions for the Weyl scalars in the metric (\ref{metr}), the
Ricci tensor components and Ricci scalar are given in Appendix
\ref{append_expr}.

\item \textbf{Conduct rotations and boosts in such a way that the new tetrad
frame be comoving with a falling observer and transform Weyl scalars and Ricci
tensor component}. Explicit transformations of the frame and corresponding
components will be given in the next Section. Corresponding rotations and
boosts are the symmetries of the Minkowski metric, so that it remains
unchanged (and thus regular), $\tilde{\eta}_{(a)(b)}=\eta_{(a)(b)}$.

\item \textbf{Deduce when Weyl scalars, Ricci tensor components and the Ricci
scalar are finite within this frame}. This will be done in Section
\ref{sec_finite}
\end{itemize}

This scheme not only allows us to find when the Riemann tensor (and thus tidal
forces) are finite but also allows us to deduce what algebraic type of
space-time an observer will see, unlike the approach in which we would
calculate the Riemann tensor by brute force.

We will be mainly interested in the vicinity of the horizon. To this end, we
consider the corresponding expansion of the metric coefficients near the
horizon
\begin{align}
A(r,\theta)=  &  A_{q}(\theta)u^{q}+o(u^{q}),~~~N^{2}(r,\theta)=\kappa
_{p}(\theta)u^{p}+o(u^{p}),\label{AN2_expan}\\
\omega=  &  \hat{\omega}_{H}+\hat{\omega}_{l}u^{l}+...+\hat{\omega}%
_{k-1}u^{k-1}+\omega_{k}(\theta)u^{k}+o(u^{k}),\label{omega_expan}\\
g_{a}=  &  g_{aH}(\theta)+g_{a1}(\theta)u+o(u),~~a=\varphi,\theta
\end{align}
where $\hat{f}$ means that corresponding quantity $f$ is independent of
$\theta$.

We assume that the Ricci scalar $R$ and other curvature scalars, such as the
Kretschmann one, have to be finite on the horizon. The conditions of their
finiteness were extensively analyzed in \cite{Ovcharenko2023}. They may be
shortly formulated in such a way.

\begin{itemize}
\item Non-extremal horizon ($p=q=1$). In this case, expansions for physical
quantities are given by:
\begin{align}
A=  &  \hat{A}_{1}u+o(u),~~~N^{2}=\hat{\kappa}_{1}%
u+o(u)\label{an_exp_non_extr}\\
\omega=  &  \hat{\omega}_{H}+\omega_{1}(\theta)u+o(u),~~~g_{a}=g_{aH}%
(\theta)+g_{a1}(\theta)u+o(u)
\end{align}

\item Extremal horizon ($q=2,~p\geq2$). In this case, expansions for physical
quantities are given by:
\begin{align}
A=  &  A_{2}(\theta)u^{2}+o(u^{2}),~~~N^{2}=\kappa_{p}(\theta)u^{p}%
+o(u^{2})\label{an_exp_extr}\\
\omega=  &  \hat{\omega}_{H}+\hat{\omega}_{1}u+...+\hat{\omega}_{k-1}%
u^{k-1}+\omega_{k}(\theta)u^{k}+o(u^{k})\\
g_{a}=  &  g_{aH}(\theta)+g_{a1}(\theta)u+o(u) \label{eh_expan}%
\end{align}
where $k\geq\left[  \dfrac{p+1}{2}\right]  .$ Hereafter, $[...]$ denotes entier.

\item Ultraextremal horizon ($q\geq3,p\geq2$). In this case, the expansions
for the metric coefficients are given by:
\begin{align}
A=  &  A_{q}(\theta)u^{q}+o(u^{2}),~~~N^{2}=\kappa_{p}(\theta)u^{p}%
+o(u^{2}),\label{an_exp_ultr_extr}\\
\omega=  &  \hat{\omega}_{H}+\hat{\omega}_{l}u^{l}+...+\hat{\omega}%
_{k-1}u^{k-1}+\omega_{k}(\theta)u^{k}+o(u^{k}),\\
g_{a}=  &  g_{aH}(\theta)+g_{a1}(\theta)u+o(u), \label{ueh_expan}%
\end{align}
where $k\geq\left[  \dfrac{p+1}{2}\right]  ,~l\geq\left[  \dfrac{p-q+3}%
{2}\right]  .$ Our notations for k, l are opposite to those in
\cite{Ovcharenko2023}.
\end{itemize}

One can check by substitution that if corresponding expansions hold, then
$\Psi_{i}$ (given by (\ref{psi_1})-(\ref{psi0})) and $\Phi_{ij}$ (given by
(\ref{phi_00})-(\ref{phi_11})) are finite. Thus we see that in the tetrad
frame under discussion the Riemann tensor and the Ricci tensors are finite if
curvature scalars are finite. This will be important for our further analysis.

Also note that knowing the behavior of the Ricci scalars allows one to deduce
how the Einstein tensor behaves, employing the relation
\begin{align}
G_{(a)(b)}=R_{(a)(b)}-\dfrac{1}{2}\eta_{(a)(b)}R.
\end{align}

As we assume the Ricci scalar to be finite, we see that the regularity of the
Einstein tensor (and thus of the energy-momentum tensor) is defined by the
regularity of the Ricci tensor that is, in its turn, defined by regularity of
Ricci scalars $\Phi_{ij}$ (\ref{phi_00})-(\ref{phi_11}).

One reservation is in order. The text of our previous paper
\cite{Ovcharenko2023} contained an inaccuracy. In the very end of Sec. 5 we
gave the values of $k$ and $l$. Actually, the corresponding values are only
the minimum ones required for regularity of the horizon. This is why after
after (\ref{eh_expan}) and (\ref{ueh_expan}) we write inequalities, not exact equalities.

\newpage

\section{Transformation to a frame, attached to a falling particle}

\label{sec_tranf} Now let us move to an analysis of the Riemann tensor in a
frame comoving with some observer that has a 4-velocity $u^{\mu}$. To obtain a
comoving frame from (\ref{ozamo_fr}), we have to rotate and boost our tetrad
vectors in such a way that in this frame $\tilde{u}^{\mu}=\{1,0,0,0\}$. We can
do this in 3 steps.

\begin{itemize}
\item Rotate a tetrad in $r\theta$ plane
\begin{align}
e_{(1)}^{\prime}=e_{(1)}\cos\psi+e_{(2)}\sin\psi\\
e_{(2)}^{\prime}=e_{(2)}\cos\psi-e_{(1)}\sin\psi
\end{align}

\item Then rotate in $r^{\prime}\varphi$ plane:
\begin{align}
e_{(1)}^{\prime\prime}  &  =e_{(1)}^{\prime}\cos\delta+e_{(3)}^{\prime}%
\sin\delta\\
e_{(3)}^{\prime\prime}  &  =e_{(3)}^{\prime}\cos\delta-e_{(1)}^{\prime}%
\sin\delta
\end{align}

\item Make a boost: The final step is to make a boost in the direction of
velocity:
\begin{align}
\tilde{e}_{(0)}  &  =\gamma(e_{(0)}^{\prime\prime}-\upsilon e_{(1)}%
^{\prime\prime})~~~~~~\tilde{e}_{(2)}=e_{(2)}^{\prime\prime}\\
\tilde{e}_{(1)}  &  =\gamma(e_{(1)}^{\prime\prime}-\upsilon e_{(0)}%
^{\prime\prime})~~~~~~\tilde{e}_{(3)}=e_{(3)}^{\prime\prime}%
\end{align}
Where $\gamma$ is a $\gamma$-factor, $V=\sqrt{1-1/\gamma^{2}}$ is absolute
value of a 3-velocity computed in a tetrad frame.
\end{itemize}

From the estimates given in Appendix \ref{append_1}, it follows that for
particles on which a finite force acts, corresponding angles $\psi$ and
$\delta$ near horizon are $=O(N)$, while $\gamma=O\left(  \dfrac{1}{N}\right)
$.

We are interested in transformations between different frames that includes
the behavior of Weyl scalars. As they are defined in terms of the null tetrad,
we need to have formulas describing this behavior directly in terms of such a
tetrad. To this end, we ask how the corresponding frame (\ref{null_fr}) is
changed under these rotations and boost and how they may be decomposed to null
rotations, spins, and boosts described in Appendix \ref{append_weyl_trans}.
This will be useful for the computation of the Weyl scalars and the Riemann
tensor components. Using Wolfram Mathematica we also showed that in principle
it is possible to rewrite transformation $e_{(a)}\rightarrow\widetilde
{e}_{(a)}$ as a combination of null rotations, spins and boost if we conduct
them in such an order:

\begin{itemize}
\item Conduct so-called spin transformation:
\[
m\rightarrow me^{i\phi_{1}},~~~k\rightarrow k,~~~l\rightarrow l
\]

\item Conduct a null rotation along direction $l$:
\[
l\rightarrow l,~~~k\rightarrow k+K\bar{m}+\bar{K}m+K\bar{K}l,~~~m\rightarrow
m+Kl
\]

\item Conduct a null rotation along direction $k$:
\[
k\rightarrow k,~~~l\rightarrow l+L\bar{m}+\bar{L}m+L\bar{L}k,~~~m\rightarrow
m+Lk
\]

\item Conduct a boost
\[
k\rightarrow Bk,~~~l\rightarrow B^{-1}l,~~~m\rightarrow m
\]

\item Conduct a spin rotation
\[
m\rightarrow me^{i\phi_{2}},~~~k\rightarrow k,~~~l\rightarrow l
\]

\end{itemize}

To match two descriptions of the transformation $e_{(a)}\rightarrow
\widetilde{e}_{(a)}$ one has to relate parameters $\varphi_{1},$ $L,$ $B,$
$K,$ $\varphi_{2}$ to $\psi,$ $\delta,$ $\gamma$ in such a way%

\begin{equation}
\tan\phi_{1}=\sin\psi\cot\delta,\text{ \ \ }\tan\phi_{2}=-\dfrac{\tan\psi
}{\sin\delta} \label{phi1_expr}%
\end{equation}

\begin{equation}
K=i\sqrt{\dfrac{1-\cos\delta\cos\psi}{1+\cos\delta\cos\psi}},\text{
\ \ }L=-\dfrac{i}{2}\sqrt{1-\cos^{2}\delta\cos^{2}\psi} \label{K_expr}%
\end{equation}

\begin{equation}
B=\dfrac{1+\cos\delta\cos\psi}{2}\dfrac{1}{\gamma+\sqrt{\gamma^{2}-1}}
\label{B_expr}%
\end{equation}

In the near-horizon limit ($N,A\rightarrow0$), as we have shown in Appendix
\ref{append_1}, $\psi,\delta\sim N$, $\gamma\sim\dfrac{1}{N}$, so that one
obtains:
\begin{equation}
\phi_{1}=\dfrac{\psi}{\delta}=-\phi_{2},~~~K=\dfrac{i}{2}\sqrt{\delta^{2}%
+\psi^{2}}=-L,~~~B=\dfrac{1}{2\gamma}%
\end{equation}
Thus in the near-horizon limit one has:
\begin{equation}
\phi_{1}=-\phi_{2}=O(1),~~~K=-L=O(N),~~~B=O(N) \label{assump_39}%
\end{equation}

\section{Conditions of finiteness of the Riemann tensor}

\label{sec_finite}

In this section we will explicitly write the regularity conditions for the
Riemann tensor in the Newman-Penrose formalism. Further, violation of some of
these conditions will enable us to build the metric of the TNBH. Using the
results from the previous Section and Appendix \ref{append_weyl_trans} (namely
(\ref{k_rot_1}-\ref{spin_6}) with $\phi_{1},$ $K,$ $B,$ $L,$ $\phi_{2}$ given
by (\ref{phi1_expr}), (\ref{K_expr}), (\ref{B_expr})), one obtains for them
(only dominant terms near the horizon are present).
\begin{align}
\tilde{\Psi}_{0}\approx &  B^{2}\Psi_{0},~~~\tilde{\Psi}_{1}\approx B\Psi
_{1},~~~\tilde{\Psi}_{2}\approx\Psi_{2},\label{psi_012}\\
\tilde{\Psi}_{3}\approx &  B^{-1}(\Psi_{3}+ke^{-i\phi_{1}}\Psi_{4}+3\bar
{L}e^{i\phi_{1}}\Psi_{2}),\label{psi_3}\\
\tilde{\Psi}_{4}\approx &  B^{-2}(\Psi_{4}+4\bar{L}e^{i\phi_{1}}\Psi_{3}%
+4\bar{L}K\Psi_{4}+6\bar{L}^{2}e^{2i\phi_{1}}\Psi_{2})\label{psi_4}\\
\tilde{\Phi}_{00}\approx &  B^{2}\Phi_{00},~~~\tilde{\Phi}_{01}\approx
B\Phi_{01},~~~\tilde{\Phi}_{11}\approx\Phi_{11},~~~\tilde{\Phi}_{02}%
\approx\Phi_{02},\\
\tilde{\Phi}_{12}\approx &  B^{-1}(\Phi_{12}+Ke^{-i\phi_{1}}\Phi
_{22}+2Le^{-i\phi_{1}}\Phi_{11}+\bar{L}e^{i\phi_{1}}\Phi_{02})\\
\tilde{\Phi}_{22}\approx &  B^{-2}[\Phi_{22}+2(Le^{-i\phi_{1}}\bar{\Phi}%
_{12}+\bar{L}e^{i\phi_{1}}\Phi_{12})+2(L\bar{K}+\bar{L}K)\Phi_{22}+\nonumber\\
&  +4L\bar{L}\Phi_{11}+(L^{2}e^{-2i\phi_{1}}\bar{\Phi}_{02}+\bar{L}%
^{2}e^{2i\phi_{1}}\Phi_{02})] \label{phi_22_boost}%
\end{align}

As according to (\ref{assump_39}) $B\sim N,~~~K\sim N$ and $\Psi_{4}=\Psi_{0}%
$, $\Psi_{3}=\Psi_{1}$, corresponding "tilded" components will be finite if
untilded ones satisfy the conditions:
\begin{equation}
\Psi_{0}=O(N^{2}),~~~\Psi_{1}=O(N),~~~\Phi_{22}=O(N^{2}),~~\Phi_{12}=O(N).
\label{main_conds}%
\end{equation}

One would obtain the same conditions if an observer was not rotating. This
follows from the fact that for non-rotating observer $\psi=\delta=0$ that,
according to (\ref{K_expr}) gives $K=L=0.$ In addition, one can check that in
this case also $B=O(N)$ (it follows from (\ref{B_expr})). Substituting this to
(\ref{psi_012}-\ref{phi_22_boost}) shows that the conditions (\ref{main_conds}%
) have to hold in this case as well. This means that adding angular momenta to
an observer does not change the conditions of the finiteness of the Riemann tensor.

Now let us analyze what exact conditions on the metric functions have to hold
for the Riemann tensor to be finite. Hereafter, we assume that coefficients in
expansions of metric functions are independent (except of coefficients in the
expansions of $N^{2}$ and $A$). Let us start with the condition for $\Psi_{0}%
$. Using (\ref{psi0}) one obtains:
\begin{align}
Re(\Psi_{0})=O(N^{2}):  &  ~\partial_{\theta}^{2}\ln(N^{2}A)-\partial_{\theta
}\ln(N^{2}A)\partial_{\theta}\ln\sqrt{\dfrac{Ag_{\varphi}g_{\theta}}{N^{2}}%
}-2\dfrac{g_{\varphi}}{N^{2}}(\partial_{\theta}\omega)^{2}+\nonumber\\
&  ~+Ag_{\theta}\left(  \partial_{r}^{2}\ln\left(  \dfrac{g_{\varphi}%
}{g_{\theta}}\right)  +\partial_{r}\ln\left(  \dfrac{g_{\varphi}}{g_{\theta}%
}\right)  \partial_{r}\ln\sqrt{\dfrac{Ag_{\varphi}g_{\theta}}{N^{2}}}\right)
=O(N^{2}),\\
Im(\Psi_{0})=O(N^{2}):  &  ~2\partial_{r}\partial_{\theta}\omega+\partial
_{r}\omega\partial_{\theta}\ln(N^{2}A)+\partial_{\theta}\omega\partial_{r}%
\ln\left(  \dfrac{g_{\varphi}^{3}}{N^{4}g_{\theta}}\right)  =O\left(
\dfrac{N^{3}}{\sqrt{A}}\right)  . \label{im_psi0_cond}%
\end{align}
The real part of $\Psi_{0}$ will be of the order of $N^{2}$ if
\begin{equation}
\partial_{\theta}\ln(N^{2}A)=O(N^{2}),~~~\partial_{\theta}\omega
=O(N^{2}),~~~\partial_{r}^{2}\ln g_{a}=O\left(  \dfrac{N^{2}}{A}\right)  ,
\label{eq_42}%
\end{equation}
where $a=\varphi,\theta$. \ It is worth noting that, according to eq. (37) of
\cite{Ovcharenko2023}, $\partial_{\theta}\omega=O(N)$. The condition
(\ref{eq_42}) on this derivative is more tight. This is not surprising since
the aforementioned equation from \cite{Ovcharenko2023} follows from the
finiteness of the Ricci scalar whereas eq. (\ref{eq_42}) is the consequence of
a more tight requirement according to which the components of the Riemann
tensor in the free falling frame should remain finite as well. We remind a
reader that it is the condition $\Psi_{0}=O(N^{2})$ that ensures the
finiteness of $\widetilde{\Psi}_{0}$ according to (\ref{psi_012}) and
(\ref{main_conds}). Note that the first condition (first derivative in
$\theta$) is sufficient, if it is fullfilled, then the second derivative will
have desired order because of (\ref{an_exp_non_extr}), (\ref{an_exp_extr}),
(\ref{an_exp_ultr_extr}). The condition for $\omega$ requires one to have an
expansion in the form given by (\ref{omega_expan}) with $k\geq p$ (we remind a
reader that finiteness of curvature scalars required $l\geq\left[
\dfrac{p-q+3}{2}\right]  ,$ see the text after (\ref{ueh_expan})).

The condition for $g_{a}$ requires $\partial_{r}\ln g_{a}=O(u^{p-q+1})$.
Integrating this condition and assuming that $g_{a}$ is regular on the horizon
(that means that cooresponding expansions starts with positive degrees of $u$)
one obtains:
\begin{equation}
g_{a}=%
\begin{cases}
g_{aH}(\theta)+g_{a,1}(\theta)u+o(u)~\mathrm{if}~p\leq q-1\\
g_{aH}(\theta)+g_{a,p-q+2}(\theta)u^{p-q+2}+o(u^{p-q+2})~\mathrm{if}~p>q-1
\end{cases}
. \label{g_a_expan}%
\end{equation}

The condition $\partial_{\theta}\ln(N^{2}A)=O(N^{2})$ is quite complicated but
it was already analyzed in Appendix C in \cite{Ovcharenko2023}, so we present
only the final result of the relation between corresponding coefficients in
the expansions of $A$ and $N^{2}$ (\ref{AN2_expan}):
\begin{align}
&  A_{q}\kappa_{p}=C_{p},~C_{p}=\mathrm{const},\nonumber\\
&  \dfrac{A_{q+s}}{A_{q}}+\dfrac{\kappa_{p+s}}{\kappa_{p}}=\sum_{n=2}%
^{s}\dfrac{(-1)^{n}}{n}\sum_{k_{j}}\dfrac{n!}{k_{1}!..k_{m}!..}\left[
\prod_{j=1}^{s}\left(  \dfrac{A_{q+j}}{A_{q}}\right)  ^{k_{j}}+\prod_{j=1}%
^{s}\left(  \dfrac{\kappa_{p+j}}{\kappa_{p}}\right)  ^{k_{j}}\right]
+C_{p+s},~\forall~s<p. \label{kappa_a_rel}%
\end{align}
In this equation, the sum is taken over all sets of $k_{j}$'s, satisfying the
condition
\begin{equation}
\sum_{j=1}^{l}jk_{j}=s.
\end{equation}

Now let us analyze $Im(\Psi_{0})=O(N^{2})$ (\ref{im_psi0_cond}). To this end,
let us at first calculate $\partial_{r}\ln\left(  \dfrac{g_{\varphi}^{3}%
}{N^{4}g_{\theta}}\right)  $:%

\begin{equation}
\partial_{r}\ln\left(  \dfrac{g_{\varphi}^{3}}{N^{4}g_{\theta}}\right)
\approx\partial_{r}\ln\left(  \dfrac{g_{\varphi H}(\theta)^{3}}{\kappa
_{p}(\theta)^{2}u^{2p}g_{\theta H}(\theta)}\right)  \approx-\dfrac{2p}{u}%
\end{equation}
Thus, analyzing the third term in (\ref{im_psi0_cond}), we see that
\begin{equation}
\partial_{\theta}\omega\partial_{r}\ln\left(  \dfrac{g_{\varphi}^{3}}%
{N^{4}g_{\theta}}\right)  \sim\dfrac{\partial_{\theta}\omega}{u}=O\left(
\dfrac{N^{3}}{\sqrt{A}}\right)
\end{equation}
that gives us the condition $\partial_{\theta}\omega=O(u^{3p/2-q/2+1})$.
Meanwhile, when we analyzed the finiteness of the real part, we obtained the
condition $\partial_{\theta}\omega\sim u^{p}$. One can see that if $p>q-2$,
this condition is stronger then $\partial_{\theta}\omega=O(u^{3p/2-q/2+1})$.
If $p\leq q-2$, the situation is opposite. Thus, employing
(\ref{an_exp_ultr_extr}-\ref{ueh_expan}), we obtain that if $p>q-2,$ $k$ has
to be greater then $p,$ while if $p\leq q-2,$ $k$ has to be greater then
$\left[  \dfrac{3p-q+3}{2}\right]  $. This number differs from $3p/2-q/2+1$ we
obtained previously because $k$ is assumed to be integer, thus we have to take
the closest integer to $3p/2-q/2+1,$ that is $\left[  \dfrac{3p-q+3}%
{2}\right]  .$

The last possible limitation on $\omega$ may appear from the second term in
(\ref{im_psi0_cond}), requiring $\partial_{r}\omega\partial_{\theta}\ln
(N^{2}A)=O\left(  \dfrac{N^{3}}{\sqrt{A}}\right)  .$ However, from eq.
(\ref{eq_42}) and finiteness of the Ricci scalar (see \cite{Ovcharenko2023})
it follows that the conditions $\partial_{\theta}(N^{2}A)=O(N^{2})$ and
$A(\partial_{r}\omega)^{2}=O(N^{2})$ hold, so the finiteness of the second
term is automatically satisfied.

To conclude, the condition $\Psi_{0}=O(N^{2})$ is satisfied if $\omega$ is
given by (\ref{omega_expan}) with $l\geq\left[  \dfrac{p-q+3}{2}\right]  $
(see \cite{Ovcharenko2023}, the end of Section V)\ with $k$ given by
\begin{equation}%
\begin{cases}
k\geq\left[  \dfrac{3p-q+3}{2}\right]  ~\mathrm{if}~p\leq q-2\\
k\geq p~\mathrm{if}~p>q-2
\end{cases}
\label{eq_48}%
\end{equation}
and $g_{a}$ is given by (\ref{g_a_expan}), and coefficients in expansions for
$A$ and $N^{2}$ are given by (\ref{kappa_a_rel}). If the aforementioned
conditions are satisfied, we have also $\Phi_{00}=O(N^{2})$.

The next condition is $\Psi_{1}=O(N)$:
\begin{align}
Re(\Psi_{1})=O(N):  &  ~2\partial_{r}\partial_{\theta}\ln\left(  \dfrac{N^{2}%
}{g_{\varphi}}\right)  +\partial_{r}\ln\left(  \dfrac{N^{2}}{g_{\varphi}%
}\right)  \partial_{\theta}\ln(N^{2}A)+\nonumber\\
&  +\partial_{\theta}\ln\left(  \dfrac{N^{2}}{g_{\varphi}}\right)
\partial_{r}\ln\dfrac{g_{\varphi}}{g_{\theta}}-4\dfrac{g_{\varphi}}{N^{2}%
}\partial_{r}\omega\partial_{\theta}\omega=O\left(  \dfrac{N}{\sqrt{A}%
}\right)  ,\label{re_psi1_cond}\\
Im(\Psi_{1})=O(N):  &  ~\partial_{\theta}^{2}\omega+\partial_{\theta}%
\omega\partial_{\theta}\ln\left(  \dfrac{g_{\varphi}^{3}A}{g_{\theta}N^{2}%
}\right)  -Ag_{\theta}\left(  \partial_{r}^{2}\omega+\frac{1}{2}\partial
_{r}\omega\partial_{r}\ln\left(  \dfrac{g_{\varphi}^{3}A}{g_{\theta}N^{2}%
}\right)  \right)  =O(N^{2}). \label{im_psi1_cond}%
\end{align}
The first term in the condition for the real part will satisfy it if
\begin{equation}
\partial_{r}\partial_{\theta}\ln N^{2}=O\left(  \dfrac{N}{\sqrt{A}}\right)  .
\label{eq_51}%
\end{equation}
Integrating this condition, one obtains $\partial_{\theta}N^{2}\sim
u^{3p/2-q/2+1}$. Using the expansion (\ref{AN2_expan}), one sees that this
requires
\begin{equation}
\kappa_{p+s}^{\prime}=0~\mathrm{for~all}~0\leq s\leq\dfrac{p-q}{2},
\label{kappa_prime_cond}%
\end{equation}
where prime here means derivative with respect to $\theta.$ Also note that if
$q>p$, this condition is absent.

Now let us consider the second term. According to (\ref{eq_42}) condition
$\partial_{\theta}\ln(N^{2}A)=O(N^{2})$ has to hold that means that the second
term (\ref{re_psi1_cond}) becomes
\begin{equation}
\partial_{r}\ln\left(  \dfrac{N^{2}}{g_{\varphi}}\right)  \sim\dfrac{1}%
{\sqrt{N^{2}A}}.
\end{equation}
Substituting corresponding expansions for $N^{2}$ and $g_{\varphi},$ we
obtain
\begin{equation}
\dfrac{\sqrt{N^{2}A}}{u}=O(1).
\end{equation}
As we assume $p$ and $q$ in the expansion of $N^{2}$ and $A$ (see eq.
(\ref{AN2_expan})) to obey $p,q\geq1$, this is indeed satisfied. So
$\partial_{\theta}\ln(N^{2}A)=O(N^{2})$ is sufficient for the second term to
be finite.

In addition, we can use the condition
\begin{equation}
\partial_{\theta}\ln(N^{2}A)=\left(  \dfrac{\partial_{\theta}N^{2}}{N^{2}%
}+\dfrac{\partial_{\theta}A}{A}\right)  =O(N^{2}) \label{an2_cond}%
\end{equation}
and eq. (\ref{eq_51}) to deduce that%
\begin{equation}
\partial_{r}\partial_{\theta}\ln A=\partial_{r}\partial_{\theta}\ln
(N^{2}A)-\partial_{r}\partial_{\theta}\ln N^{2}%
\end{equation}

The first term in RHS of this equation is $O(N^{2}),$ the second is $O\left(
\dfrac{N}{\sqrt{A}}\right)  .$ As we assume $p,q>0,$ it is obvious that
$N^{2}$ is of higher order than $\dfrac{N}{\sqrt{A}}.$ This means that in
dominant order we can write%
\begin{equation}
\partial_{r}\partial_{\theta}\ln A=O\left(  \dfrac{N}{\sqrt{A}}\right)  .
\end{equation}
\ 

Resolving this we obtain
\begin{equation}
A_{q+s}^{\prime}=0~\mathrm{for~all}~0\leq s\leq\dfrac{p-q}{2}.
\label{a_prime_cond}%
\end{equation}

Also note that if $q>p$, this condition is absent.

If the condition (\ref{kappa_prime_cond}) holds, then the third term in
(\ref{re_psi1_cond}) also satisfies the regularity condition.

The last requirement comes from the fourth term in (\ref{re_psi1_cond}).
However, as $\partial_{\theta}\omega\sim N^{2}$ (see eq. (\ref{eq_42})) and
$\partial_{r}\omega\sim\dfrac{N}{\sqrt{A}}$ (the condition of the finiteness
of curvature scalars, see \cite{Ovcharenko2023}), we see that this term also
satisfies the regularity condition.

Now let us move to the requirement that comes from the imaginary part of the
$\Psi_{1}$ (\ref{im_psi1_cond}). The first two terms already satisfy it
because $\partial_{\theta}\omega=O(N^{2})$, while the third and the fourth
terms are more complicated. They will satify corresponding requirements in the
right hand side of (\ref{im_psi1_cond}) if $\partial_{r}\omega\sim u^{p-q+1}$.
(It is also assumed that on the horizon $\dfrac{g_{\varphi H}^{3}A_{q}%
}{g_{\theta H}\kappa_{p}^{2}}$ depends on $\theta,$ we will not focus on a
very specific case when this is not so). However, it was obtained previously
\cite{Ovcharenko2023} that the Ricci scalar is finite if the condition
$\partial_{r}\omega=O(N/\sqrt{A})\sim u^{(p-q)/2}$ holds (eq. (39) in
\cite{Ovcharenko2023}). This condition is stronger then $\partial_{r}%
\omega\sim u^{p-q+1}$ if $p\leq q-2$. Otherwise, the $\partial_{r}\omega\sim
u^{p-q+1}$ is dominant.

Thus the requirement $\Psi_{1}=O(N)$ is satisfied if eq.
(\ref{kappa_a_expan_2}) is fulfilled.%

\begin{equation}%
\begin{cases}
\kappa_{p+s}^{\prime}=A_{p+s}^{\prime}=0~\mathrm{for}~0\leq s\leq\dfrac
{p-q}{2},\\
\mathrm{Relation}~(\ref{kappa_a_rel})~\mathrm{holds~for~}\dfrac{p-q}{2}<s<p,\\
\mathrm{No~special~condition~for~}s\geq p.
\end{cases}
\label{kappa_a_expan_2}%
\end{equation}

Additionally, the number $k$ in the expansion for $\omega$ (\ref{omega_expan})
should obey $k\geq\left[  \dfrac{p+1}{2}\right]  $ according to what is said
below eqs. (\ref{eh_expan}) and (\ref{ueh_expan}). As far as $l$ is concerned,
we have
\begin{equation}%
\begin{cases}
l\geq\left[  \dfrac{p-q+3}{2}\right]  ~\mathrm{if}~p\leq q-2,\\
l\geq p-q+2~\mathrm{if}~p>q-2.
\end{cases}
\label{eq_60}%
\end{equation}

Expressions for $l$ are obtained by integrating expressions $\partial
_{r}\omega\sim u^{p-q+1}$ and $\partial_{r}\omega\sim u^{(p-q)/2}$ found
above. Here, we introduced a square bracket notation to stress that we are
considering only integer $k$ and $l$. For example, if $p-q$ is an odd number,
we have to take the closest higher integer number to $\frac{p-q}{2}+1$. Taking
a closest higher value is required to be sure that the regularity conditions
are satisfied, that is the entier of $\dfrac{p-q+3}{2}.$ Thus, to denote
ceiling of number $\frac{p-q}{2}+1$ we use a notation $\left[  \dfrac
{p-q+3}{2}\right]  .$

If relations (\ref{kappa_a_expan_2}) hold, one can check that $\Phi_{01}=O(N)$
(\ref{phi_01}) is also satisfied.

Also, let us formulate the conditions that have to hold for both $\Psi
_{0}=O(N^{2})$ and $\Psi_{1}=O(N)$ to be satisfied. In this case, $g_{a}$ has
to satisfy (\ref{g_a_expan}), coefficients in expansions for $A$ and $N^{2}$
have to satisfy (\ref{kappa_a_expan_2}) and expansion for $\omega$ has to be
given by (\ref{omega_expan}) with $k$ and $l$ satisfying the relation (this
may be obtained by combining (\ref{eq_48}) and (\ref{eq_60})).
\begin{equation}%
\begin{cases}
k\geq\left[  \dfrac{3p-q+3}{2}\right]  ,~l\geq\left[  \dfrac{p-q+3}{2}\right]
~\mathrm{if}~p\leq q-2,\\
k\geq p,~l\geq p-q+2~\mathrm{if}~p>q-2.
\end{cases}
\end{equation}
In this case also one can check that $\Phi_{00}=O(N^{2})$ and $\Phi_{01}=O(N)$
are satified.

Also one can check that these relations are exactly the same as the ones
obtained in \cite{Ovcharenko2023} when we were analyzing the regularity
conditions. However, current analysis is required because we do it in the
Newman-Penrose formalism that will also be usefil further for algebraic classifications.

It may be instructive to write them in a more compact form. The horizon is
regular if the expansion for the metric functions is such (here we introduced
a new quantity $m$ which represents a degree of the dominant term in expansion
for $g_{a}$):
\begin{align}
N^{2}=  &  \kappa_{p}(\theta)u^{p}+o(u^{p}),~A=A_{q}(\theta)u^{q}%
+o(u^{q}),\label{n2a_expan_2}\\
\omega=  &  \hat{\omega}_{H}+\hat{\omega}_{l}u^{l}+...+\omega_{k}(\theta
)u^{k}+o(u^{k}),\label{omega_expan_2}\\
g_{a}=  &  g_{aH}(\theta)+g_{am}(\theta)u^{m}+o(m), \label{ga_expan_2}%
\end{align}
with the coefficients in expansion of $A$ and $N^{2}$ satisfying the relations
what follow from combining conditions (\ref{kappa_a_rel},
\ref{kappa_prime_cond}, \ref{a_prime_cond})%
\begin{equation}%
\begin{cases}
\kappa_{p+s}^{\prime}=A_{p+s}^{\prime}=0~\mathrm{for}~0\leq s\leq s_{1},\\
\mathrm{Relation}~(\ref{kappa_a_rel})~\mathrm{holds~for~}s_{1}<s<s_{2},\\
\mathrm{No~special~condition~for~}s\geq s_{2},
\end{cases}
\label{kappa_a_expan_3}%
\end{equation}
where
\begin{align}
l\geq &  \max\left(  \left[  \dfrac{p-q+3}{2}\right]  ,p-q+2\right)
,~~~k\geq\max\left(  \left[  \dfrac{3p-q+3}{2}\right]  ,p\right)  ,\nonumber\\
m  &  \geq\max(1,p-q+2),~~~s_{1}=\dfrac{p-q}{2},~~~s_{2}=p.
\label{numbers_regular}%
\end{align}

In addition, one may ask what relations have to be held to satisfy $\Psi
_{0}=O(N^{2})$ and $\Psi_{1}=O(N)$ separately. From the text above it follows
that
\begin{align}
\Psi_{0}=O(N^{2}):~~~l\geq &  \left[  \dfrac{p-q+3}{2}\right]  ,~~~k\geq
\max\left(  p,\left[  \dfrac{3p-q+3}{2}\right]  \right)  ,\nonumber\\
s_{1}=  &  0,~~~s_{2}=p,~~~m\geq\max(1,p-q+2)\\
\Psi_{1}=O(N):~~~l\geq &  \max\left(  \left[  \dfrac{p-q+3}{2}\right]
,p-q+2\right)  ,~~~k\geq\left[  \dfrac{p+1}{2}\right]  ,\nonumber\\
s_{1}=  &  \dfrac{p-q}{2},~~~s_{2}=p,~~~m\geq1.
\end{align}

\section{Behavior of Weyl scalars and classification of horizons}

The main aim of our work is to suggest generalization of the notion "naked
horizon" and give corresponding classification that generalizes \cite{nk1}
\ \cite{Zaslavskii2007} for rotating metrics (\ref{metr}). (We remind a reader
that in the present work we restrict ourselves by 3+1 space-times (\ref{metr})
and do not consider higher dimensions \cite{nk1}). This classification is
based on account of behavior of the components of the curvature tensor in a
free-falling frame. It was observed in \cite{nk1} that in principle, after
rotations and boost, there may be one (or several) components of the Riemann
tensor that are significantly enhanced in the free-falling frame as compared
to the static one. (Hereafter, we will use the term FZAMO, if a frame is
attached to a geodesic observer with the zero angular momentum, F stands for
"free falling". Although we assume that reference particles are moving freely,
the frame and corresonding results can be generalized to the case of nonzero
finite acceleration. In the horizon limit FZAMO and OZAMO (\ref{ozamo_fr})
behave essentially different since, in general, the acceleration of reference
particle diverge.) The existence of such horizons was demonstrated for
dilatonic, $U(1)^{2}$ gravity and several exact solutions from the string
theory \cite{nk1}, \cite{nk2}. The next step was made in \cite{pravda2005}
where it was noticed that some components of the curvature tensor in the FZAMO
frame can be not simply large but even formally diverging. How this phenomenon
reveals itself in the spherically symmetric and distorted static metrics was
considered in \cite{Zaslavskii2007}. In the latter work, the author introduced
such a definition: if at least one from components of the Riemann tensor
diverges, then the horizon is truly naked. If all these components tend to
zero, then the horizon is usual, while if all the components are finite
(except of the case when all they tend to zero), then such a horizon is naked.

If the metric is static and spherically symmetric, there is only one relevant
quantity $Z$ in terms of which this definition was done. If it is static but
does not possess spherical symmetry, there are three relevant quantities
$Z_{ab}$ (where $a,b=\theta$,$\phi)$ - see \cite{Zaslavskii2007} for details.
In our case situation is more complicated (see below). As it follows from
(\ref{psi_012})-(\ref{phi_22_boost}), after a boost components $\tilde{\Psi
}_{0}$, $\tilde{\Psi}_{1}$, $\tilde{\Phi}_{00}$, $\tilde{\Phi}_{01}$ tend to
zero, components $\tilde{\Psi}_{2}$, $\tilde{\Phi}_{11}$ $\tilde{\Phi}_{02}$
remain finite (as untilded are finite) and $\tilde{\Psi}_{3}$, $\tilde{\Psi
}_{4}$, $\tilde{\Phi}_{12}$, $\tilde{\Phi}_{22}$ are potentially divergent (we
have to note that now we make a further step and allow conditions from the
previous Section to be violated that means that now some quantities
potentially diverge). Thus modifications of the previous analysis are required.

\begin{itemize}
\item Horizon is \textbf{usual} if all potentially divergent scalars
$\tilde{\Psi}_{4},~\tilde{\Psi}_{3},~\tilde{\Phi}_{22},~\tilde{\Phi}_{12}$
after rotations and boost \textbf{tend to zero}. This equivalently means that
$\Psi_{0}=o(N^{2}),~\Psi_{1}=o(N),~\Phi_{00}=o(N^{2}),~\Phi_{01}=o(N)$.

\item Horizon is \textbf{naked} if at least one of the potentially divergent
scalars $\tilde{\Psi}_{4},~\tilde{\Psi}_{3},~\tilde{\Phi}_{22},~\tilde{\Phi
}_{12}$ \textbf{is finite and separated from zero} (while all others may tend
to zero)

\item Horizon is \textbf{truly naked} if at least one of the potentially
divergent scalars $\tilde{\Psi}_{4},~\tilde{\Psi}_{3},~\tilde{\Phi}%
_{22},~\tilde{\Phi}_{12}$ \textbf{diverges} (while all others may be finite or
tend to zero)
\end{itemize}

In the previous section, we investigated the regularity conditions, which are
very useful in formulating conditions when the horizon is usual, naked, or
truly naked. Horizon is usual if the expansions (\ref{n2a_expan_2}%
)-(\ref{ga_expan_2}) hold with conditions (\ref{kappa_a_expan_3}). However, if
we want potentially divergent scalars to tend to zero, all the numbers
$l,~k,~m,~s_{1},~s_{2}$ have to be greater then the values, listed in
(\ref{numbers_regular}):
\begin{align}
l>  &  \max\left(  \left[  \dfrac{p-q+3}{2}\right]  ,p-q+2\right)
,~~~k>\max\left(  \left[  \dfrac{3p-q+3}{2}\right]  ,p\right)  ,\nonumber\\
m  &  >\max(1,p-q+1),~~~s_{1}>\dfrac{p-q}{2},~~~s_{2}>p. \label{numbers_usual}%
\end{align}

If the horizon is naked, then at least one (but not necessarily all) of the
numbers $l,~k,~m,~s_{1},~s_{2}$ have to be equal to a value, listed in
(\ref{numbers_regular}), while all others may be greater or equal to these values.

If a horizon is truly naked, then at least one of the numbers has to be lower
than the ones listed in (\ref{numbers_regular}).

The corresponding set of all conditions when the spacetime is usual, naked, or
truly naked are listed in Tab. \ref{tab1}. In addition, one may be interested
when corresponding components are finite distinctly as we represent in Tab.
\ref{tab1_1}.

\begin{table}[ptb]
\centering
\begin{tabular}
[c]{|c||c|}\hline
Horizon & Conditions\\\hline\hline
Usual & $l>\max\Big(\Big[\dfrac{p-q+3}{2}\Big],p-q+2\Big)~\mathrm{\wedge
}~k>\max\Big(\Big[\dfrac{3p-q+3}{2}\Big],p\Big)$\\
~ & $\wedge$ $m>\max(1,p-q+2)~\mathrm{\wedge}~s_{1}>\dfrac{p-q}{2}
~\mathrm{\wedge}~s_{2}>p.$\\\hline
Naked & $l\geq\max\Big(\Big[\dfrac{p-q+3}{2}\Big],p-q+2\Big)~\mathrm{\wedge
}~k\geq\max\Big(\Big[\dfrac{3p-q+3}{2}\Big],p\Big)$\\
~ & $\wedge$ $m\geq\max(1,p-q+2)~\mathrm{\wedge}~s_{1}\geq\dfrac{p-q}
{2}~\mathrm{\wedge}~s_{2}\geq p.$\\
~ & except of conditions for usual horizon\\\hline
Truly naked & $0<l<\max\Big(\Big[\dfrac{p-q+3}{2}\Big],p-q+2\Big)$ $\vee$
$0<k<\max\Big(\Big[\dfrac{3p-q+3}{2}\Big],p\Big)$\\
~ & $\vee$ $0<m<\max(1,p-q+2)$ $\vee$ $0<s_{1}<\dfrac{p-q}{2}$ $\vee$
$0<s_{2}<p$\\\hline
\end{tabular}
\caption{Table showing conditions that have to hold for black hole to be
usual, naked and truly naked.}%
\label{tab1}%
\end{table}

\begin{table}[ptb]
\centering
\begin{tabular}
[c]{|c||c|}\hline
Near-horizon behaviour & Conditions\\\hline\hline
$\Psi_{0}=o(N^{2})$, $\tilde{\Psi}_{0}$ tends to zero & $l\geq\Big[\dfrac
{p-q+3}{2}\Big]$ $\wedge$ $k>\max\Big(p,\Big[\dfrac{3p-q+3}{2}\Big]\Big)$\\
~ & $\wedge$ $s_{1}\geq0$ $\wedge$ $s_{2}>p$ $\wedge$ $m>\max(1,p-q+2)$%
\\\hline
$\Psi_{0}=O(N^{2})$, $\tilde{\Psi}_{0}$ is finite and non-zero &
$l\geq\Big[\dfrac{p-q+3}{2}\Big]$ $\wedge$ $k\geq\max\Big(p,\Big[\dfrac
{3p-q+3}{2}\Big]\Big)$\\
~ & $\wedge$ $s_{1}\geq0$ $\wedge$ $s_{2}\geq p$ $\wedge$ $m\geq\max
(1,p-q+2)$\\
~ & except of conditions for $\Psi_{0}=o(N^{2})$\\\hline
$\Psi_{0}\neq O(N^{2})$, $\tilde{\Psi}_{0}$ diverges & $0<l<\Big[\dfrac
{p-q+3}{2}\Big]$ $\vee$ $0<k<\max\Big(p,\Big[\dfrac{3p-q+3}{2}\Big]\Big)$\\
~ & $\vee$ $0<s_{2}<p$ $\vee$ $0<m<\max(1,p-q+2)$\\\hline
$\Psi_{1}=o(N)$, $\tilde{\Psi}_{1}$ tends to zero & $l>\max\Big(\Big[\dfrac
{p-q+3}{2}\Big],p-q+2\Big)$ $\wedge$ $k\geq\Big[\dfrac{p+1}{2}\Big]$\\
~ & $\wedge$ $s_{1}>\dfrac{p-q}{2}$ $\wedge$ $s_{2}>p$\\\hline
$\Psi_{1}=O(N)$, $\tilde{\Psi}_{1}$ is finite and non-zero & $l\geq
\max\Big(\Big[\dfrac{p-q+3}{2}\Big],p-q+2\Big)$ $\wedge$ $k\geq\Big[\dfrac
{p+1}{2}\Big]$\\
~ & $\wedge$ $s_{1}\geq\dfrac{p-q}{2}$ $\wedge$ $s_{2}\geq p$\\
~ & except of conditions for $\Psi_{1}=o(N)$, $\tilde{\Psi}_{1}$
diverges\\\hline
$\Psi_{1}\neq O(N)$ & $0<l<\max\Big(\Big[\dfrac{p-q+3}{2}\Big],p-q+2\Big)$
$\vee$ $0<k<\Big[\dfrac{p+1}{2}\Big]$\\
~ & $\vee$ $0<s_{1}<\dfrac{p-q}{2}$ $\vee$ $0<s_{2}<p$.\\\hline
\end{tabular}
\caption{Table showing a set of conditions when $\widetilde{\Psi}_{0}$ and
$\widetilde{\Psi}_{1}$ tend to zero, are finite but separated from zero, or
infinite. Notation $\neq O$ means that the corresponding quantity tends to
zero with the lower rate then $N$ or $N^{2}.$}%
\label{tab1_1}%
\end{table}

\section{Relation to algebraic types}

Now let us investigate the relation of the algebraic type of considered
space-times and the regularity conditions. For simplicity, let us focus on the
vacuum solutions (this means that $R_{\mu\nu}=0$, and, thus, $\Phi_{ij}=0$).
However, before we proceed further, we have to be careful with what we call an
algebraic type, because there are different approaches to the algebraic
classification. If a space-time contains a black hole, one should distinguish
between the algebraic type away from the horizon (called "off-horizon"
\cite{Tanatarov2014}) and that on the horizon itself ("on-horizon" type). If
one asks what algebraic type the OZAMO will see near the horizon, then such a
type is called "boosted type". In addition, one may ask what algebraic type
will be seen by FZAMO. This is the "regular type". Our aim is to find which
conditions have to hold for each of these types for a horizon to be regular.
Let us start with the off-horizon type. As we deal with the axially symmetric
stationary spacetime, this imposes several additional conditions and
simplifications on the algebraic classification. Following the results
formulated in \cite{Tanatarov2014} (see, especially, Sec. V there), one will
obtain these off-horizon types if such conditions hold (also note that our
notations differ from the ones, listed in \cite{Tanatarov2014}):

\begin{itemize}
\item Type O: $\Psi_{0}=\Psi_{1}=\Psi_{2}=0$

\item Type N: $\Psi_{0}=-\Psi_{2}=\pm i \Psi_{1}$

\item Type III: impossible for axially symmetric stationary spacetimes

\item Type D: either $\Psi_{1}=0$ and $3\Psi_{2}-\Psi_{0}=0$ or $2 \Psi
_{1}^{2}=\Psi_{0}(3\Psi_{2}-\Psi_{0})$

\item Type II: $\pm4 i \Psi_{1}=3\Psi_{2}-\Psi_{0}$
\end{itemize}

The boosted type is defined as the algebraic type, calculated from quantities
$\Psi_{i}$ near the horizon. If the horizon is regular, then, as we have
shown, conditions $\Psi_{0}=O(N^{2}),~\Psi_{1}=O(N)$ have to hold, so in the
near horizon limit they both tend to zero. In addition, due to axial symmetry
$\Psi_{4}=\Psi_{0}$, $\Psi_{3}=\Psi_{1}$, so these Weyl scalars also tend to
zero. Thus, we are left with Weyl scalar is $\Psi_{2}$ only, which can be
either zero or non-zero. In the first case, one obtains algebraic type O, in
the second-type D, so that there are only these 2 boosted types.

The regular type is defined by the Weyl scalars $\tilde{\Psi}_{i}$, obtained
after rotations and boost. As follows from (\ref{psi_012}), for regular
spacetimes $\tilde{\Psi}_{0},~\tilde{\Psi}_{1}\rightarrow0$, so that regular
types are only II, D, III, N, O, with type I excluded.

Let us start with the off-horizon type O. In this case, $\Psi_{0}=\Psi
_{1}=\Psi_{2} $, so that we see, that the Weyl, and thus the Riemann, tensors
are zero. For such spacetimes no special conditions are required. Also, note
that the boosted and regular types are O.

If the spacetime is of off-horizon type N, then
\begin{equation}
\Psi_{0}=\Psi_{2}=\pm i\Psi_{1}. \label{type_n_cond}%
\end{equation}
Regularity requires $\Psi_{0}=O(N^{2})$ (\ref{main_conds}). From
(\ref{type_n_cond}) it follows that in this case $\Psi_{2}=O(N^{2})$ and
$\Psi_{1}=O(N^{2})$, so in the near-horizon limit all Weyl scalars tend to
zero. This means that the boosted type is O. To find the regular type we have
to use (\ref{psi_012})-(\ref{psi_4}) and conditions $\Psi_{0}\sim\Psi_{1}%
\sim\Psi_{2}\sim N^{2}$. From this it follows that the only possible non-zero
"tilded" component is $\tilde{\Psi}_{4}$, so this spacetime is of regular type
N (unless the exceptional case when $\Psi_{4}$ is zero, then the regular type is O).

If the spacetime is of off-horizon type D, then there are 2 possibilities:
either $\Psi_{1}=0$ and $3\Psi_{2}-\Psi_{0}=0$ or $2\Psi_{1}^{2}=\Psi
_{0}(3\Psi_{2}-\Psi_{0})$. In the first case $\Psi_{1}$ is automatically zero,
so that condition $\Psi_{1}=O(N)$ is automatically satisfied. From condition
$3\Psi_{2}-\Psi_{0}=0$ and regularity condition $\Psi_{0}=O(N^{2})$
(\ref{main_conds}) we find $\Psi_{2}=O(N^{2})$, so all Weyl scalars near the
horizon tend to zero. This means that the boosted type is O, while the regular
type is N. In the second case we see that it is enough to have the condition
$\Psi_{0}=O(N^{2})$ satisfied: in this case from the condition $2\Psi_{1}%
^{2}=\Psi_{0}(3\Psi_{2}-\Psi_{0})$ we automatically have $\Psi_{2}=O(1)$,
$\Psi_{1}=O(N)$. Thus the boosted type is D, while the regular could be in
principle II. However, note that after the boost the condition $2\widetilde
{\Psi}_{1}^{2}=\widetilde{\Psi}_{0}(3\widetilde{\Psi}_{2}-\widetilde{\Psi}%
_{0})$ is also satified in a boosted components, so that the regular type is
also D.

The last case is the off-horizon type II. In this case both conditions
$\Psi_{1}=O(N)$ and $\Psi_{0}=O(N^{2})$ have to be satisfied separately. This
is so because the defining condition of the algebraic type II, namely
$\pm4i\Psi_{1}=3\Psi_{2}-\Psi_{0}$, does not give the condition $\Psi
_{1}=O(N)$ as a consequence of the condition $\Psi_{0}=O(N^{2})$ (and vise
versa). However, this condition can be inverted and gives $\Psi_{2}=O(N)$.
From this follows that the boosted type is O, while the regular one is III.
This follows from the fact that in this case $\Psi_{0}\sim N^{2}$, $\Psi
_{1}\sim N$, $\Psi_{2}\sim N$ and the transformations (\ref{psi_012}%
-\ref{phi_22_boost}) that give $\tilde{\Psi}_{0}\sim N^{2} \Psi_{0}\sim N^{4}%
$, $\tilde{\Psi}_{1}\sim N \Psi_{1}\sim N^{2}$, $\tilde{\Psi}_{2}\sim N$,
$\tilde{\Psi}_{3}\sim\Psi_{3}/N=O(1)$, $\tilde{\Psi}_{4}\sim\Psi_{4}%
/N^{2}=O(1)$.

We summarize all results in Table \ref{tab_1}. In this Table we, along with
the results for naked horizons (which we presented above) also present results
for the usual and truly naked horizons, which can be easily derived by
imposing $\Psi_{0}=o(N^{2})$ or $\Psi_{1}=o(N)$ instead of regularity
conditions (usual horizons), or violating these conditions (truly naked
horizons). Note that the relation between the off-horizon and on-horizon
algebraic types is the same as the one obtained in \cite{Tanatarov2014} (in
this work the authors imposed regularity conditions from the beginning, so
that the case of TNBHs, violating them, could not be considered there in principle).

\begin{table}[ptb]
\centering%
\begin{tabular}
[c]{|c||c|c|c|c|}\hline
Type of horizon & Off-horizon type PT & Regular PT & Boosted PT & Additional
condition\\\hline\hline
Usual & O & O & O & $-$\\\hline
Usual & N & O & O & $\Psi_{0}=o(N^{2})$\\
Naked & ~ & N & O & $\Psi_{0}=O(N^{2})$\\
Truly naked & ~ & III & O & $\Psi_{0}\neq O(N^{2})$\\\hline
Usual & D, $\Psi_{1}=0$ & O & O & $\Psi_{0}=o(N^{2})$\\
Naked & ~ & N & O & $\Psi_{0}=O(N^{2})$\\
Truly naked & ~ & N & O & $\Psi_{0}\neq O(N^{2})$\\\hline
Usual & D, $\Psi_{1}\neq0$ & D & D & $\Psi_{0}=o(N^{2})$\\
Naked & ~ & D & D & $\Psi_{0}=O(N^{2})$\\
Truly naked & ~ & II & D & $\Psi_{0}\neq O(N^{2})$\\\hline
Usual & II & III & O & $\Psi_{0}=o(N^{2}),~\Psi_{1}=o(N)$\\
Naked & ~ & III & O & $\Psi_{0}=O(N^{2})~\mathrm{or}~\Psi_{1}=O(N)$\\
Truly naked & ~ & III & O & $\Psi_{0}\neq O(N^{2})~\mathrm{or}~\Psi_{1}\neq
O(N)$\\\hline
\end{tabular}
\caption{Relations between off-horizon Petrov type, regular Petrov type,
boosted Petrov type and regularity conditions which for each of them hold to
have usual, naked or truly naked horizon (here we assume that the spacetime is
vacuum). PT in the name of columns stands for \textquotedblleft Petrov type".}%
\label{tab_1}%
\end{table}

\section{Static spherically symmetric case}

To check validity of our approach, in this Section we compare the above resuts
with those obtained earlier for the static spherically symmetric case.

The corresponding metric can be written in the form%

\begin{equation}
ds^{2}=-N^{2}dt^{2}+\frac{dr^{2}}{A}+r^{2}(d\theta^{2}+\sin^{2}\theta
d\varphi^{2}), \label{metss}%
\end{equation}

where $N^{2}=N(r)^{2},$ $A=A(r).$ This metric can be obtained from
(\ref{metr}) if
\begin{equation}
g_{\theta\theta}=r^{2},\text{ \ \ }g_{\varphi\varphi}=r^{2}\sin^{2}%
\theta,\text{ \ \ }\omega=0. \label{stat_siml}%
\end{equation}

Using our most general expressions for Weyl scalars (\ref{psi_1}-\ref{psi0})
we note that if we substitute in these expressions simplified metric functions
(\ref{stat_siml}), we will obtain $\Psi_{0}=\Psi_{4}=0$ and $\Psi_{1}=\Psi
_{3}=0$. This is expected because it is known that spherically symmetric
spacetimes are of algebraic type D (see, e.g. \cite{wald}, p. 187). Moreover,
this simplification gives us $\Phi_{01}=-\overline{\Phi}_{12}=0.$ The only
non-zero potentially divergent Ricci tensor components read near the horizon%
\begin{equation}
\Phi_{22}=\Phi_{00}\approx\frac{A_{q}}{4}\frac{p-q}{r_{h}}u^{q-1}.
\end{equation}

Thus we see that in the static limit there is only one potentially-divergent
quantity. After a boost we will have%
\begin{equation}
\widetilde{\Phi}_{22}\sim\frac{\Phi_{22}}{N^{2}}\sim(p-q)u^{q-1-p}.
\end{equation}

Meanwhile, for a spherically symmetric metric $\Phi_{22}$ coincides with the
quantity $Z$ introduced in eq. (8) of \cite{Zaslavskii2007}. Moreover, our
definition of TNBHs, naked and usual black holes (whether $\tilde{\Psi}_{3}$,
$\tilde{\Psi}_{4}$, $\tilde{\Phi}_{12}$, $\tilde{\Phi}_{22}$ are divergent,
finite, or tend to zero) in the static limit reduces to the definitions given
in \cite{Zaslavskii2007} (whether $\widetilde{Z}$ is divergent, finite, or
tend to zero, where $\widetilde{Z}$ is the quantity $Z$ after boost). This is
quite natural because, as we have already shown, for static spherically
symmetric space-times $\Psi_{3}=\Psi_{4}=0,$ $\Phi_{12}=0$ and thus there is
only one quantity, required for description of potentially divergent
components, $\Phi_{22}=Z$.

\section{Conclusions}

In this work we built classification of horizons that generalizes the notion
of naked black holes \cite{nk1} and TNBHs \cite{pravda2005},
\cite{Zaslavskii2007} to the rotating case. To this end, we utilized the
Neman-Penrose formalism. Unlike the static case, this definition requires
introduction of 4 quantities, which have a direct geometrical interpretation.
These are the boosted Weyl scalars $\tilde{\Psi}_{4},~\tilde{\Psi}_{3},$ and
the boosted Ricci scalars $\tilde{\Phi}_{22},~\tilde{\Phi}_{12}.$ We have
shown that conditions of their finiteness agree with ones obtained in
\cite{Ovcharenko2023}. What is more interesting, we have developed conditions
when these quantities either tend to zero (what corresponds to usual black
holes) or they are non-zero but finite (naked black holes) or diverge (truly
naked black holes). Physically, we obtained classification based on quantities
that generalize simple picture of tidal forces near the horizon (for usual and
naked black holes) or its analogue in the case of TNBHs. Corresponding results
are formulated in terms of numbers that characterize the behavior of the
metric coeficients near the horizon. These results are presented in Table
\ref{tab1}. From the mathematical viewpoint, our results for TNBH describe
properties of nonscalar polynomial light-like singularities as counterparts of
black holes.

Usage of the Newman-Penrose formalism allowed us to analyze how conditions for
a black hole to be usual, naked or truly naked affect their algebraic type
that is collected in Table \ref{tab_1}, where we, according to
\cite{pravda2005} and \cite{Tanatarov2014}, distinguish between off-horizon
and on-horizon algebraic types. Our results for the algebraic types of usual
and naked black holes agree with \cite{pravda2005} and \cite{Tanatarov2014}.
We also found algebraic types for TNBHs that was not considered previously.
Also, we have verified our approach in the case of static spacetime and obtain
consistency with previous works.

\appendix

\section{Expressions for the Weyl scalars, Ricci tensor and Ricci scalar}

\label{append_expr}

\textbf{Weyl scalars}
\begin{align}
\Psi_{1}=  &  \dfrac{-1}{16}\sqrt{\dfrac{A}{g_{\theta}}}\left\{  2\partial
_{r}\partial_{\theta}\ln\left(  \dfrac{N^{2}}{g_{\varphi}}\right)
+\partial_{r}\ln\left(  \dfrac{N^{2}}{g_{\varphi}}\right)  \partial_{\theta
}\ln(N^{2}A)+\right. \nonumber\\
&  \left.  +\partial_{\theta}\ln\left(  \dfrac{N^{2}}{g_{\varphi}}\right)
\partial_{r}\ln\dfrac{g_{\varphi}}{g_{\theta}}-4\dfrac{g_{\varphi}}{N^{2}%
}\partial_{r}\omega\partial_{\theta}\omega\right\} \nonumber\\
&  -\dfrac{i}{8g_{\theta}}\sqrt{\dfrac{g_{\varphi}}{N^{2}}}\left[
\partial_{\theta}^{2}\omega+\frac{1}{2}\partial_{\theta}\omega\partial
_{\theta}\ln\left(  \dfrac{g_{\varphi}^{3}A}{g_{\theta}N^{2}}\right)
-Ag_{\theta}\left(  \partial_{r}^{2}\omega+\frac{1}{2}\partial_{r}%
\omega\partial_{r}\ln\left(  \dfrac{g_{\varphi}^{3}A}{g_{\theta}N^{2}}\right)
\right)  \right] \label{psi_1}\\
3\Psi_{2}  &  -\Psi_{0}=\dfrac{-1}{4g_{\theta}}\left\{  \partial_{\theta}%
^{2}\ln\left(  \dfrac{N^{2}}{g_{\varphi}}\right)  +\dfrac{1}{2}\partial
_{\theta}\ln\left(  \dfrac{N^{2}}{g_{\varphi}}\right)  \partial_{\theta}%
\ln\left(  N^{2}A\dfrac{g_{\varphi}}{g_{\theta}}\right)  -2\dfrac{g_{\varphi}%
}{N^{2}}(\partial_{\theta}\omega)^{2}-\right. \nonumber\\
&  ~~~\left.  -Ag_{\theta}\left(  \partial_{r}^{2}\ln\left(  \dfrac{N^{2}%
}{g_{\varphi}}\right)  +\dfrac{1}{2}\partial_{r}\ln\left(  \dfrac{N^{2}%
}{g_{\varphi}}\right)  \partial_{r}\ln\left(  N^{2}A\dfrac{g_{\varphi}%
}{g_{\theta}}\right)  -2\dfrac{g_{\varphi}}{N^{2}}(\partial_{r}\omega
)^{2}\right)  \right\} \nonumber\\
&  ~~~+\dfrac{i}{2}\sqrt{\dfrac{Ag_{\varphi}}{N^{2}g_{\theta}}}\left[
2\partial_{r}\partial_{\theta}\omega+\frac{1}{2}\partial_{\theta}%
\omega\partial_{r}\ln\left(  \dfrac{g_{\varphi}^{3}}{N^{2}g_{\theta}^{2}%
}\right)  +\frac{1}{2}\partial_{r}\omega\partial_{\theta}\ln\left(
\dfrac{g_{\varphi}^{3}A^{2}}{N^{2}}\right)  \right] \label{3psi2-psi0}\\
\Psi_{0}=  &  \dfrac{1}{8g_{\theta\theta}}\left\{  \partial_{\theta}^{2}%
\ln(N^{2}A)-\frac{1}{2}\partial_{\theta}\ln(N^{2}A)\partial_{\theta}\ln\left(
\dfrac{Ag_{\varphi}g_{\theta}}{N^{2}}\right)  -2\dfrac{g_{\varphi}}{N^{2}%
}(\partial_{\theta}\omega)^{2}+\right. \nonumber\\
&  \left.  +Ag_{\theta}\left(  \partial_{r}^{2}\ln\left(  \dfrac{g_{\varphi}%
}{g_{\theta}}\right)  +\frac{1}{2}\partial_{r}\ln\left(  \dfrac{g_{\varphi}%
}{g_{\theta}}\right)  \partial_{r}\ln\left(  \dfrac{Ag_{\varphi}g_{\theta}%
}{N^{2}}\right)  \right)  \right\} \nonumber\\
&  -\dfrac{i}{8}\sqrt{\dfrac{Ag_{\varphi}}{N^{2}g_{\theta}}}\left[
2\partial_{r}\partial_{\theta}\omega+\partial_{r}\omega\partial_{\theta}%
\ln(N^{2}A)+\partial_{\theta}\omega\partial_{r}\ln\left(  \dfrac{g_{\varphi
}^{3}}{N^{4}g_{\theta}}\right)  \right] \label{psi0}\\
\Psi_{4}=  &  \Psi_{0},~~~\Psi_{3}=-\Psi_{1}%
\end{align}
Please note that the original metric is invariant with respect to inversions
of the time and angular coordinates
\begin{equation}
t\rightarrow-t,~~~\varphi\rightarrow-\varphi
\end{equation}
what yields such a change in the null tetrad:
\begin{equation}
k\rightarrow-l,~~~l\rightarrow-k,~~m\rightarrow\bar{m}%
\end{equation}
From the condition that the Weyl tensor is invariant with respect to these
transformations, one obtains that
\begin{equation}
\Psi_{4}=\Psi_{0},~~~\Psi_{3}=-\Psi_{1}%
\end{equation}

\textbf{Ricci tensor components}:
\begin{align}
\Phi_{00}=  &  \dfrac{1}{8g_{\theta}}\left\{  \partial_{\theta}^{2}\ln
(N^{2}A)+\dfrac{1}{2}\partial_{\theta}\ln(N^{2}A)\partial_{\theta}\ln\left(
\frac{N^{2}g_{\varphi}}{Ag_{\theta}}\right)  -g_{\varphi}\dfrac{(\partial
_{\theta}\omega)^{2}}{N^{2}}-\right. \nonumber\\
&  \left.  -Ag_{\theta}\left[  \partial_{r}^{2}\ln(g_{\varphi}g_{\theta
})+\dfrac{1}{4}\partial_{r}\ln(g_{\varphi}g_{\theta})\partial_{r}\ln\left(
\dfrac{g_{\varphi}g_{\theta}A^{2}}{N^{4}}\right)  +\dfrac{1}{4}\left(
\partial_{r}\ln\dfrac{g_{\varphi}}{g_{\theta}}\right)  ^{2}\right]  \right\}
\label{phi_00}\\
\Phi_{01}=  &  \dfrac{-1}{16}\sqrt{\dfrac{A}{g_{\theta}}}\Bigg\{2\partial
_{r}\partial_{\theta}\ln(N^{2}g_{\varphi})+\partial_{r}\ln(N^{2}g_{\varphi
})\partial_{\theta}\ln(N^{2}A)-\nonumber\\
&  -\partial_{\theta}\ln(N^{2}g_{\varphi})\partial_{r}\ln g_{\theta}%
-\partial_{\theta}\ln\left(  \dfrac{N^{2}}{g_{\varphi}}\right)  \partial
_{r}\ln g_{\varphi}-2g_{\varphi}\dfrac{\partial_{r}\omega\partial_{\theta
}\omega}{N^{2}}-\nonumber\\
&  \left.  -2\dfrac{i}{g_{\varphi}}\left[  \partial_{r}\left(  \sqrt
{\dfrac{Ag_{\theta}g_{\varphi}^{3}}{N^{2}}}\partial_{r}\omega\right)
+\partial_{\theta}\left(  \sqrt{\dfrac{g_{\varphi}^{3}}{g_{\theta}AN^{2}}%
}\partial_{\theta}\omega\right)  \right]  \right\} \label{phi_01}\\
\Phi_{02}=  &  \dfrac{1}{8g_{\theta\theta}}\left\{  \partial_{\theta}^{2}%
\ln\left(  \dfrac{A}{N^{2}}\right)  +\dfrac{1}{4}\partial_{\theta}\ln\left(
\dfrac{A}{N^{2}}\right)  \partial_{\theta}\ln\left(  \dfrac{N^{2}}%
{Ag_{\varphi}^{2}g_{\theta}^{2}}\right)  -\dfrac{1}{4}(\partial_{\theta}%
\ln(N^{2}A))^{2}+\right. \nonumber\\
&  \left.  2\dfrac{g_{\varphi}}{N^{2}}(\partial_{\theta}\omega)^{2}%
+Ag_{\theta}\left[  \partial_{r}^{2}\ln\left(  \dfrac{g_{\varphi}}{g_{\theta}%
}\right)  +\dfrac{1}{2}\partial_{r}\ln\left(  \frac{g_{\varphi}}{g_{\theta}%
}\right)  \partial_{r}\ln(N^{2}Ag_{\varphi}g_{\theta})+\dfrac{g_{\varphi}%
}{N^{2}}(\partial_{r}\omega)^{2}\right]  \right\} \label{phi_02}\\
\Phi_{11}=  &  \dfrac{-1}{8g_{\theta}}\left\{  \partial_{\theta}^{2}\ln
g_{\varphi}+\dfrac{1}{2}\partial_{\theta}\ln g_{\varphi}\partial_{\theta}%
\ln\left(  \dfrac{g_{\varphi}}{g_{\theta}}\right)  +\dfrac{1}{8}\left[
\left(  \partial_{\theta}\ln(N^{2}A)\right)  ^{2}-\left(  \partial_{\theta}%
\ln\left(  \dfrac{N^{2}}{A}\right)  \right)  ^{2}\right]  -\right. \nonumber\\
&  -Ag_{\theta}\left[  \partial_{r}^{2}\ln N^{2}+\dfrac{1}{2}\partial_{r}\ln
N^{2}\partial_{r}\ln(N^{2}A)-\dfrac{1}{8}\left(  (\partial_{r}\ln g_{\theta
}g_{\varphi})^{2}-\left(  \partial_{r}\ln\left(  \dfrac{g_{\theta}}%
{g_{\varphi}}\right)  \right)  ^{2}\right)  \right] \nonumber\\
&  \left.  +\frac{g_{\varphi}}{2N^{2}}\left(  (\partial_{\theta}\omega
)^{2}+3Ag_{\theta}(\partial_{r}\omega)^{2}\right)  \right\} \label{phi_11}\\
\Phi_{22}=  &  \Phi_{00},~~~\Phi_{12}=-\bar{\Phi}_{01}%
\end{align}
\textbf{Ricci scalar}
\begin{align}
R=  &  \dfrac{-1}{g_{\theta}}\left\{  \partial_{\theta}^{2}\ln\dfrac{N^{2}}%
{A}+\dfrac{1}{2}\partial_{\theta}\ln\dfrac{N^{2}}{A}\partial_{\theta}\ln
\dfrac{N^{2}g_{\varphi}}{Ag_{\theta}}+\dfrac{1}{8}\left[  \left(
\partial_{\theta}\ln(N^{2}A)\right)  ^{2}-\left(  \partial_{\theta}\ln\left(
\dfrac{N^{2}}{A}\right)  \right)  ^{2}\right]  +\right. \nonumber\\
&  +\partial_{\theta}^{2}\ln g_{\varphi}+\dfrac{1}{2}\partial_{\theta}\ln
g_{\varphi}\partial_{\theta}\ln\dfrac{g_{\varphi}}{g_{\theta}}+\dfrac
{g_{\varphi}}{2N^{2}}(\partial_{\theta}\omega)^{2}-Ag_{\theta}\Bigg[\partial
_{r}^{2}\ln N^{2}+\nonumber\\
&  +\dfrac{1}{2}\partial_{r}\ln N^{2}\partial_{r}\ln(N^{2}A)+\partial_{r}%
^{2}\ln(g_{\varphi}g_{\theta})+\dfrac{1}{2}\partial_{r}\ln(g_{\varphi
}g_{\theta})\partial_{r}\ln(N^{2}Ag_{\varphi}g_{\theta})-\nonumber\\
&  \left.  \left.  -\dfrac{1}{8}\left(  \left(  \partial_{r}\ln(g_{\varphi
}g_{\theta})\right)  ^{2}-\left(  \partial_{r}\ln\dfrac{g_{\varphi}}%
{g_{\theta}}\right)  ^{2}\right)  -2\dfrac{g_{\varphi}}{N^{2}}(\partial
_{r}\omega)^{2}\right]  \right\}  \label{R_expr}%
\end{align}

All these expressions were obtained in Wolfram Mathematica 13.3

\section{The behavior of the velocity near horizon}

\label{append_1}

In this section, we discuss the behavior of the 4-velocity near horizon
relevant in our context. As the metric is invariant with respect to $t$ and
$\varphi$ translations, corresponding conservation laws give us:
\begin{equation}
u^{t}=\dfrac{X}{N^{2}},~\mathrm{where}~X=\mathcal{E}-\omega\mathcal{L}%
,~~~u^{\varphi}=\dfrac{\mathcal{L}}{g_{\varphi\varphi}}+\dfrac{\omega X}%
{N^{2}},
\end{equation}
where $\mathcal{E}$ and $\mathcal{L}$ are the specific (per unit mass) energy
and the component of the angular momentum generated by rotation in $\phi$
direction. Normalization for 4-velocity $u^{\mu}u_{\mu}=1$ entails:
\begin{equation}
u^{r}=\sigma\sqrt{A}\dfrac{\sqrt{X^{2}-N^{2}(1+\mathcal{L}^{2}/g_{\varphi
}+g_{\theta}(u^{\theta})^{2})}}{N}.
\end{equation}

Here $\sigma$ is a sign showing direction of motion. Hereafter,we consider
only "usual" particles (without fine-tuning of parameters).

The component $u^{\theta}$ of the four-velocity can be defined from the
geodesics equation but for our analysis it will be sufficient to take a
natural assumption that $u^{\theta}$ is finite near horizon. This means that
$u^{r}\sim\dfrac{\sqrt{A}}{N}$ near horizon.

Now let us analyze behavior of a trajectory near the horizon in the OZAMO
frame. To do this, first of all, we have to compute the components of
3-velocity, defined by relation:
\begin{equation}
V^{(i)}=\dfrac{e_{\mu}^{(i)}u^{\mu}}{e_{\mu}^{(0)}u^{\mu}}.
\end{equation}

Using (\ref{ozamo_fr}), we can get:
\begin{gather}
V^{(1)}=\dfrac{u^{r}}{X}\dfrac{N}{\sqrt{A}},~~~V^{(2)}=\sqrt{g_{\theta}}%
\dfrac{u^{\theta}N}{X},\\
V^{(3)}=\dfrac{\mathcal{L}N}{\sqrt{g_{\varphi}}X}.
\end{gather}

Angles in the $r\theta$ and $r\varphi$ planes are defined as
\begin{equation}
\tan\psi=\dfrac{V^{(2)}}{V^{(1)}}\sim O(N),~~~~~~\tan\delta=\dfrac{V^{(3)}%
}{V^{(1)}}\sim O(N).
\end{equation}

So, both angles $\sim O(N).$ In addition, let us calculate the square of the
3-velocity:
\begin{equation}
V^{2}=(V^{(1)})^{2}+(V^{(2)})^{2}+(V^{(3)})^{2}=1-\dfrac{N^{2}}{X^{2}}%
\end{equation}
Thus the gamma factor is
\begin{equation}
\gamma=\dfrac{X}{N}.
\end{equation}
so that near the horizon $\gamma\sim\dfrac{1}{N}$

\section{Riemann tensor in OZAMO frame}

The only non-trivial components of the Riemann tensor are (here we denote
$R_{abcd}^{\prime}=R_{(a)(b)(c)(d)}$)%

\begin{align}
R_{0101}^{\prime}  &  =\frac{A}{4}\left[  2\partial_{r}^{2}\ln N^{2}%
+\partial_{r}\ln N^{2}\partial_{r}\ln(AN^{2})\frac{{}}{{}}\right. \\
&  \left.  -\frac{1}{Ag_{\theta\theta}}\left(  \partial_{\theta}\ln
N^{2}\partial_{\theta}\ln A+3\frac{A}{N^{2}}g_{\varphi\varphi}g_{\theta\theta
}(\partial_{r}\omega)^{2}\right)  \right] \\
R_{0102}^{\prime}  &  =\frac{1}{4}\sqrt{\frac{A}{g_{\varphi\varphi}}}\left[
2\partial_{r}\partial_{\theta}\ln N^{2}+\partial_{r}\ln N^{2}\partial_{\theta
}\ln(AN^{2})-\partial_{\theta}\ln N^{2}\partial_{r}\ln g_{\theta\theta
}-3g_{\varphi\varphi}\frac{\partial_{\theta}\omega\partial_{r}\omega}{N^{2}%
}\right] \\
R_{0113}^{\prime}  &  =-\frac{A}{4}\sqrt{\frac{g_{\varphi\varphi}}{N^{2}}%
}\left[  2\partial_{r}^{2}\omega+\partial_{r}\omega\partial_{r}\ln\left(
\frac{g_{\varphi\varphi}^{3}A}{N^{2}}\right)  -\frac{1}{Ag_{\theta\theta}%
}\partial_{\theta}\omega\partial_{\theta}\ln A\right] \\
R_{0123}^{\prime}  &  =-\frac{1}{4}\sqrt{\frac{Ag_{\varphi\varphi}}%
{N^{2}g_{\theta\theta}}}\left[  2\partial_{r}\partial_{\theta}\omega
+\partial_{r}\omega\partial_{\theta}\ln\left(  \frac{Ag_{\varphi\varphi}^{2}%
}{N^{2}}\right)  +\partial_{\theta}\omega\partial_{r}\ln\left(  \frac
{g_{\varphi\varphi}}{g_{\theta\theta}}\right)  \right] \\
R_{0202}^{\prime}  &  =\frac{1}{4g_{\theta\theta}}\left[  2\partial_{\theta
}^{2}\ln N^{2}+\partial_{\theta}\ln N^{2}\partial_{\theta}\ln\left(
\frac{N^{2}}{g_{\theta\theta}}\right)  -3g_{\varphi\varphi}\frac
{(\partial_{\theta}\omega)^{2}}{N^{2}}+Ag_{\theta\theta}\partial_{r}\ln
N^{2}\partial_{r}\ln g_{\theta\theta}\right] \\
R_{0213}^{\prime}  &  =-\frac{1}{4}\sqrt{\frac{Ag_{\varphi\varphi}}%
{N^{2}g_{\theta\theta}}}\left[  2\partial_{r}\partial_{\theta}\omega
+\partial_{r}\omega\partial_{\theta}\ln\left(  Ag_{\varphi\varphi}\right)
+\partial_{\theta}\omega\partial_{r}\ln\left(  \frac{g_{\varphi\varphi}^{2}%
}{N^{2}g_{\theta\theta}}\right)  \right] \\
R_{0223}^{\prime}  &  =-\frac{1}{4g_{\theta\theta}}\sqrt{\frac{g_{\varphi
\varphi}}{N^{2}}}\left[  2\partial_{\theta}^{2}\omega+\partial_{\theta}%
\omega\partial_{\theta}\ln\left(  \frac{g_{\varphi\varphi}^{3}}{N^{2}%
g_{\theta\theta}}\right)  +Ag_{\theta\theta}\partial_{r}\omega\partial_{r}\ln
g_{\theta\theta}\right] \\
R_{0303}^{\prime}  &  =\frac{1}{4g_{\theta\theta}}\left[  \partial_{\theta}\ln
N^{2}\partial_{\theta}\ln g_{\varphi\varphi}+Ag_{\theta\theta}\partial_{r}\ln
N^{2}\partial_{r}\ln g_{\varphi\varphi}+\frac{(\partial_{\theta}\omega
)^{2}+Ag_{\theta\theta}(\partial_{r}\omega)^{2}}{N^{2}}\right] \\
R_{0312}^{\prime}  &  =\frac{1}{4}\sqrt{\frac{Ag_{\varphi\varphi}}%
{N^{2}g_{\theta\theta}}}\left[  \partial_{\theta}\omega\partial_{r}\ln\left(
\frac{N^{2}}{g_{\varphi\varphi}}\right)  -\partial_{r}\omega\partial_{\theta
}\ln\left(  \frac{N^{2}}{g_{\varphi\varphi}}\right)  \right] \\
R_{1212}^{\prime}  &  =\frac{1}{4g_{\theta\theta}}\left[  \left(
2\partial_{\theta}^{2}\ln A-\partial_{\theta}\ln A\partial_{\theta}%
\ln(Ag_{\theta\theta})\right)  -Ag_{\theta\theta}(2\partial_{r}^{2}\ln
g_{\theta\theta}+\partial_{r}\ln g_{\theta\theta}\partial_{r}\ln
(Ag_{\theta\theta}))\right] \\
R_{1313}^{\prime}  &  =-\frac{A}{4}\left[  2\partial_{r}^{2}\ln g_{\varphi
\varphi}+\partial_{r}\ln g_{\varphi\varphi}\partial_{r}\ln(Ag_{\varphi\varphi
})-\frac{1}{Ag_{\varphi\varphi}}\partial_{\theta}\ln A\partial_{\theta}\ln
g_{\varphi\varphi}+\frac{g_{\varphi\varphi}}{N^{2}}(\partial_{r}\omega
)^{2}\right] \\
R_{1332}^{\prime}  &  =\frac{1}{4}\sqrt{\frac{A}{g_{\theta\theta}}}\left[
2\partial_{r}\partial_{\theta}\ln g_{\varphi\varphi}+\partial_{r}\ln
g_{\varphi\varphi}\partial_{\theta}\ln(Ag_{\varphi\varphi})-\partial_{\theta
}\ln g_{\varphi\varphi}\partial_{r}\ln g_{\theta\theta}+g_{\varphi\varphi
}\frac{\partial_{\theta}\omega\partial_{r}\omega}{N^{2}}\right]
\end{align}

With respect to a boost these components transform in such a way (here
$\tilde{R}_{abcd}$ are boosted components of the Riemann tensor).%

\begin{gather}
\tilde{R}_{0202}=R_{0202}^{\prime}~~~~~~\tilde{R}_{0213}=R_{0213}^{\prime
}~~~~~\tilde{R}_{1313}=R_{1313}^{\prime}\\
\tilde{R}_{0101}=R_{0101}^{\prime}+\sinh^{2}\gamma(R_{0101}^{\prime}%
+R_{1212}^{\prime}),\text{ \ \ }\tilde{R}_{2323}=R_{2323}^{\prime}+\sinh
^{2}\gamma(R_{2323}^{\prime}+R_{0303}^{\prime})\\
\tilde{R}_{0303}=R_{0303}^{\prime}+\sinh^{2}{\gamma}(R_{0303}^{\prime
}+R_{2323}^{\prime}),\text{ \ \ }\tilde{R}_{1212}=R_{1212}^{\prime}+\sinh
^{2}{\gamma}(R_{1212}^{\prime}+R_{1010}^{\prime})\\
\tilde{R}_{0312}=R_{0312}^{\prime}+\sinh^{2}\gamma(R_{0312}^{\prime}%
+R_{0123}^{\prime}),\text{ \ \ }\tilde{R}_{0123}=R_{0123}^{\prime}+\sinh
^{2}\gamma(R_{0123}^{\prime}-R_{0312}^{\prime})\\
\tilde{R}_{1223}=\sinh\gamma\cos\gamma(R_{0123}^{\prime}-R_{0312}^{\prime
}),\text{ \ \ }\tilde{R}_{0103}=\sinh\gamma\cosh\gamma(R_{0123}^{\prime
}-R_{0312}^{\prime})\\
\tilde{R}_{0121}=\cosh2\gamma R_{0121}^{\prime}+\cosh\gamma\sinh
\gamma(R_{0101}^{\prime}+R_{1212}^{\prime})\\
\tilde{R}_{0323}=\cosh2\gamma R_{0323}^{\prime}+\cosh\gamma\sinh
\gamma(R_{0303}^{\prime}+R_{2323}^{\prime})\\
\tilde{R}_{0113}=\cosh\gamma R_{0113}^{\prime},\text{ }\tilde{R}_{0223}%
=\cosh\gamma R_{0223}^{\prime},\text{ }\tilde{R}_{1323}=\cosh\gamma
R_{1323}^{\prime},\text{ }\tilde{R}_{0201}=\cosh\gamma R_{0201}^{\prime}\\
\tilde{R}_{0203}=\sinh\gamma R_{0223}^{\prime},\text{ }\tilde{R}_{1213}%
=-\sinh\gamma R_{0113}^{\prime},\text{ }\tilde{R}_{0212}=-\sinh\gamma
R_{0102}^{\prime},\text{ }\tilde{R}_{0313}=\sinh\gamma R_{1323}^{\prime}%
\end{gather}

\section{Transformation of the Weyl scalars and Ricci tensor components}

\label{append_weyl_trans} In this Appendix, we list how Weyl scalars and Ricci
tensor components change under the action of different symmetries of the null
tetrad frame.

\textbf{$k$-rotations}
\begin{align}
k^{\prime}=  &  k,~~~m^{\prime}=m+Lk,~~~\bar{m}^{\prime}=\bar{m}+\bar
{L}k,~~~l^{\prime}=l+L\bar{m}+\bar{L}m+L\bar{L}k,\label{k_rot_1}\\
\Psi_{0}^{\prime}=  &  \Psi_{0},\\
\Psi_{1}^{\prime}=  &  \Psi_{1}+\bar{L}\Psi_{0},~~~\\
\Psi_{2}^{\prime}=  &  \Psi_{2}+2\bar{L}\Psi_{1}+\bar{L}^{2}\Psi_{0},~~~\\
\Psi_{3}^{\prime}=  &  \Psi_{3}+3\bar{L}\Psi_{2}+3\bar{L}^{2}\Psi_{1}+\bar
{L}^{3}\Psi_{0},\\
\Psi_{4}^{\prime}  &  =\Psi_{4}+4\bar{L}\Psi_{3}+6\bar{L}^{2}\Psi_{2}+4\bar
{L}^{3}\Psi_{1}+\bar{L}^{4}\Psi_{4}\\
\Phi_{00}^{\prime}  &  =\Phi_{00}\\
\Phi_{01}^{\prime}  &  =\Phi_{01}+L\Phi_{00}\\
\Phi_{02}^{\prime}  &  =\Phi_{02}+2L\Phi_{01}+L^{2}\Phi_{00},\\
\Phi_{11}^{\prime}  &  =\Phi_{11}+L\bar{\Phi_{01}}+\bar{L}\Phi_{01}+L\bar
{L}\Phi_{00}\\
\Phi_{12}^{\prime}=  &  \Phi_{12}+2L\Phi_{11}+\bar{L}\Phi_{02}+L^{2}\bar{\Phi
}_{01}+2L\bar{L}\Phi_{01}+L^{2}\bar{L}\Phi_{00}\nonumber\\
\Phi_{22}^{\prime}=  &  \Phi_{22}+2(L\bar{\Phi}_{12}+\bar{L}\Phi_{12}%
)+4L\bar{L}\Phi_{11}+L^{2}\bar{\Phi}_{02}+\bar{L}^{2}\Phi_{02}+2L\bar{L}%
(L\bar{\Phi}_{01}+\bar{L}\Phi_{01})+(L\bar{L})^{2}\Phi_{00} \label{k_rot_8}%
\end{align}
\textbf{$l$-rotations}
\begin{align}
l^{\prime}=  &  l,~~~m^{\prime}=m+Kl,~~~\bar{m}^{\prime}=\bar{m}+\bar
{K}l,~~~k^{\prime}=k+K\bar{m}+\bar{K}m+K\bar{K}l,\label{l_rot_1}\\
\Psi_{4}^{\prime}=  &  \Psi_{4},~~~\\
\Psi_{3}^{\prime}=  &  \Psi_{3}+K\Psi_{4},~~~\\
\Psi_{2}^{\prime}=  &  \Psi_{2}+2K\Psi_{3}+K^{2}\Psi_{4},~~\\
\Psi_{1}^{\prime}=  &  \Psi_{1}+3K\Psi_{2}+3K^{2}\Psi_{3}+K^{3}\Psi_{4},\\
\Psi_{0}^{\prime}  &  =\Psi_{0}+4K\Psi_{1}+6K^{2}\Psi_{2}+4K^{3}\Psi_{3}%
+K^{4}\Psi_{4}\\
\Phi_{22}^{\prime}  &  =\Phi_{22}\\
\Phi_{12}^{\prime}  &  =\Phi_{12}+K\Phi_{22}\\
~\Phi_{02}^{\prime}  &  =\Phi_{02}+2K\Phi_{12}+K^{2}\Phi_{22},\\
\Phi_{11}^{\prime}  &  =\Phi_{11}+K\bar{\Phi_{12}}+\bar{K}\Phi_{12}+K\bar
{K}\Phi_{22}\\
\Phi_{01}^{\prime}=  &  \Phi_{01}+2K\Phi_{11}+\bar{K}\Phi_{02}+K^{2}\bar{\Phi
}_{12}+2K\bar{K}\Phi_{12}+K^{2}\bar{K}\Phi_{22}\\
\Phi_{00}^{\prime}=  &  \Phi_{00}+2(K\bar{\Phi}_{01}+\bar{K}\Phi_{01}%
)+4K\bar{K}\Phi_{11}+K^{2}\bar{\Phi}_{02}+\bar{K}^{2}\Phi_{02}+\nonumber\\
&  +2K\bar{K}(K\bar{\Phi}_{12}+\bar{K}\Phi_{12})+(K\bar{K})^{2}\Phi_{22}
\label{l_rot_9}%
\end{align}
\textbf{Boosts}
\begin{align}
k^{\prime}=  &  Bk,~~~l^{\prime}=B^{-1}l,~~~m^{\prime}=m,~~~\bar{m}^{\prime
}=\bar{m}\label{boost_1}\\
\Psi_{0}^{\prime}=  &  B^{2}\Psi_{0},~~\Psi_{1}^{\prime}=B\Psi_{1}%
,~~~~\Psi_{2}^{\prime}=\Psi_{2},\\
\Psi_{3}^{\prime}=  &  B^{-1}\Psi_{3},~~\Psi_{4}^{\prime}=B^{-2}\Psi_{4},~~~\\
\Phi_{00}^{\prime}  &  =B^{2}\Phi_{00},\text{ \ \ }\Phi_{01}^{\prime}%
=B\Phi_{01},\text{ \ \ }\Phi_{02}^{\prime}=\Phi_{02},\\
\Phi_{11}^{\prime}  &  =\Phi_{11},\text{ \ \ }~\Phi_{12}^{\prime}=B^{-1}%
\Phi_{12},\text{ \ \ }\Phi_{22}^{\prime}=B^{-2}\Phi_{22}. \label{boost_6}%
\end{align}
\textbf{Spin rotations}
\begin{align}
k^{\prime}=  &  k,~~~l^{\prime}=l,~~~m^{\prime}=e^{i\theta}m,~~~\bar
{m}^{\prime}=e^{-i\theta}\bar{m}\label{spin_1}\\
\Psi_{0}^{\prime}=  &  e^{2i\theta}\Psi_{0},~~~\Psi_{1}^{\prime}=e^{i\theta
}\Psi_{1},\text{ \ \ }\Psi_{2}^{\prime}=\Psi_{2},~~\\
\Psi_{3}^{\prime}=  &  e^{-i\theta}\Psi_{3},\text{ \ \ }\Psi_{4}^{\prime
}=e^{-2i\theta}\Psi_{4}.\\
\Phi_{02}^{\prime}  &  =e^{2i\theta}\Phi_{02},\text{ \ \ }\Phi_{01}^{\prime
}=e^{i\theta}\Phi_{01},\text{ \ \ }\Phi_{12}^{\prime}=e^{i\theta}\Phi_{12},\\
\Phi_{11}^{\prime}  &  =\Phi_{11},\text{ \ \ }\Phi_{22}^{\prime}=\Phi
_{22},\text{ \ \ }\Phi_{00}^{\prime}=\Phi_{00}, \label{spin_6}%
\end{align}

\end{document}